\def\ion#1#2{#1$\;${\small\rm\@Roman{#2}}\relax}
\documentclass[bibyear]{aa}

\usepackage{graphicx}
\usepackage{txfonts}

\def\go{
\mathrel{\raise.3ex\hbox{$>$}\mkern-14mu\lower0.6ex\hbox{$\sim$}}
}
\def\lo{
\mathrel{\raise.3ex\hbox{$<$}\mkern-14mu\lower0.6ex\hbox{$\sim$}}
}
\begin{document}
   \title{X-ray and UV observations of V751~Cyg in an optical high state}
   \author{K.L. Page \and J.P. Osborne \and A.P. Beardmore \and P.A. Evans \and S.R. Rosen \and M.G. Watson 
          }

   \offprints{klp5@leicester.ac.uk}

   \institute{Dept. of Physics and Astronomy, University of Leicester, Leicester, LE1 7RH, UK
}              

   \date{Received ; accepted }

  \abstract{}{The VY~Scl system (anti-dwarf nova) V751~Cyg is examined following a claim of a
super-soft spectrum in the optical low state.}
	    {A serendipitous \textit{XMM-Newton} X-ray observation and, 21 months later, \textit{Swift} X-ray and UV observations, have provided the best such data on this source so
far.  These optical high-state datasets are used to study the flux and spectral variability of V751~Cyg.}
	      {Both the \textit{XMM-Newton} and \textit{Swift} data show evidence for modulation of the X-rays for the first time at the known 3.467~hr orbital period of V751~Cyg. In two \textit{Swift} observations, taken ten days apart, the mean X-ray flux remained unchanged, while the UV source brightened by half a magnitude. The X-ray spectrum was not super-soft during the optical high state, but rather due to multi-temperature optically thin emission, with significant (10$^{21-22}$~cm$^{-2}$) absorption, which was higher in the observation by \textit{Swift} than that of \textit{XMM-Newton}. The X-ray flux is harder at orbital minimum, suggesting that the modulation is related to absorption, perhaps linked to the azimuthally asymmetric wind absorption seen previously in H$\alpha$.}
	      {}
   

   \titlerunning{V751 Cyg}
   \authorrunning{K.L. Page et al.} 
  \maketitle
%

\section{Introduction}

\label{intro}

During the pre-release verification of the 3XMM-DR4 X-ray source
catalogue (Rosen et al., in prep), examination of a sample of bright
($f_X$~$>$~10$^{-12}$~erg~cm$^{-2}$~s$^{-1}$) point sources, variable
on short timescales, revealed a serendipitous detection of the
cataclysmic variable V751 Cyg. Previously reported as a transient
super-soft X-ray source (Greiner et al. 1999), the relatively hard {\em XMM-Newton}
spectrum invited further investigation. The 3XMM-DR4 catalogue
contains over a quarter of a million unique X-ray sources found in
13 years of {\em XMM-Newton} observations; for the brighter sources,
time series and spectra are provided. The ease with
which this interesting observation of V751 Cyg was found is a
testament to the value of the catalogue.

V751~Cyg is classed as a nova-like variable and VY~Scl (anti-dwarf nova) system (Ritter \& Kolb 2003). The term `nova-like variable' includes all non-eruptive cataclysmic variables (CVs) for which the accretion disc is believed to be permanently in the high-viscosity state associated with dwarf nova outbursts. The VY Scl stars are a small subset of these which occasionally undergo a drop in flux to a low state (Leach et al. 1999, Honeycutt \& Kafka 2004). These low states can last for several hundred days and are believed to be caused by intervals of low mass-transfer in the system.

V751~Cyg was previously observed in the X-ray band (Greiner et
al. 1999) by the {\em ROSAT}-PSPC (Position Sensitive Proportional
Counter) and HRI (High Resolution Imager). PSPC observations occurred
in 1990 and 1992 when the optical source was in its usual high state,
providing only upper limits for the source X-ray count rate, while the
HRI data collected in June and December 1997 (during which time the
optical emission was in the low state) resulted in a detection of
V751~Cyg. Greiner (1998) reports that the X-ray luminosity determined
using the PSPC was below 2--6~$\times$~10$^{30}$ erg~s$^{-1}$ during the
optical high state, while, during the optical low state, Greiner et
al. (1999) find that it had a super-soft blackbody-like spectrum with
kT~=~15$^{+15}_{-10}$~eV and a luminosity in the range 7~$\times$10$^{33-36}$ erg~s$^{-1}$ (for a distance of 500~pc), despite the HRI having a variable
gain and very modest spectral resolution (FWHM/E = 0.66-1.07, Fraser
1992\footnote{Microchannel Plate Energy Discrimination: ftp://legacy.gsfc.nasa.gov/rosat/doc/hri/spec\_resp\_zero.txt}); this led them to suggest an anti-correlation of X-ray and
optical intensity, as was seen in RX~J0513.9$-$6951 (e.g. Reinsch et
al. 1996).


Patterson et al. (2001) measured the orbital period of V751~Cyg to be 0.1445(2) day, with a most likely value of 0.144464(1)~day (3.467~hr), using radial velocities; a modulation had also been previously noted in optical data by Walker \& Bell (1980), though they appear only to give the time interval over which the modulation was observed, rather than the actual period of that modulation.  Patterson et al. (2001) also discuss the difficulties in
accommodating a high luminosity central source in the optical low state.

In this paper, we present serendipitious {\em XMM-Newton} observations, along with follow-up {\em Swift} data (all obtained when V751~Cyg was in its optical high state), investigating the timing (Section \ref{timinganal}) and spectral (Section~\ref{specanal}) properties of the source, using both X-ray and UV data.

\section{Observations}
\label{obs}

\begin{figure}
\centering
\includegraphics[width=8cm]{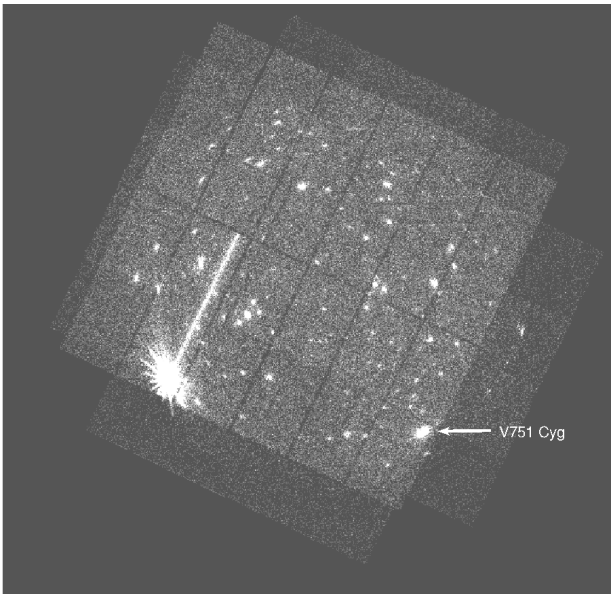}
\caption{The {\em XMM-Newton} image of observation 0679580201, with V751~Cyg marked. The bright source towards the bottom left is the chromospherically active, rapidly rotating, late-type giant star HD 199178 (V1794 Cyg; e.g., Gondoin 2004). North is up, East to the left and the rotation of the image is due to the roll angle of the spacecraft ($\sim$~244$^o$). The image is plotted using a log intensity scale.}
\label{fov}
\end{figure}

V751~Cyg was serendipitously observed by {\em XMM-Newton} (Jansen et al. 2001) during an observation of NGC~7000 (the North America Nebula) on 2011 Nov. 24 (Table~\ref{table:obs}); the field of view is shown in Fig.~\ref{fov}. The Optical Monitor was blocked, due to bright stars in the vicinity. The 3XMM-DR4 catalogue source name corresponding to V751 Cyg is 3XMM~J205212.7+441925:482810; this X-ray source, which has a position error of 0.4~arcsec (1$\sigma$), lies within 0.8~arcsec of the 2MASS All Sky Catalog position of V751~Cyg. The exposure time of the observation was $\sim$~26~ks in pn and $\sim$~29~ks for the MOS instruments.


\begin{figure}
\centering
\includegraphics[width=6cm, angle=-90]{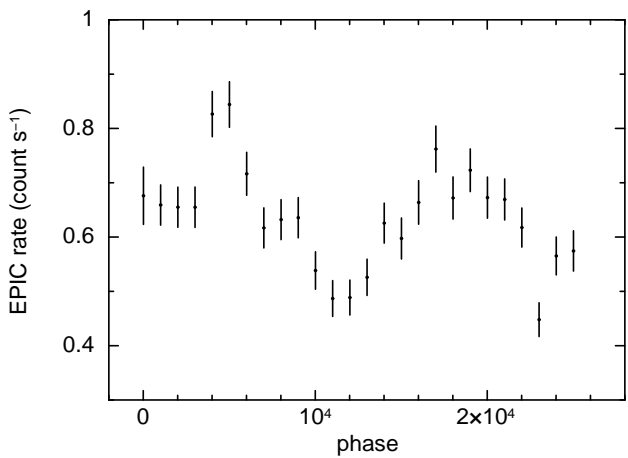}
\caption{1000s-binned {\em XMM-Newton} combined EPIC (0.3--10~keV) X-ray light-curve of V751~Cyg, showing possible modulation at the orbital period of 3.467~hr ($\sim$~12.5~ks). The time origin of the plot is seconds since the time from which all three instruments were taking data: 2011 Nov. 24 at 10:32 = MJD 55889.44 (see Table~\ref{table:obs}).}
\label{lc}
\end{figure}

The {\em XMM-Newton} X-ray light-curve (Fig.~\ref{lc}) revealed two peaks separated by approximately the period given by Patterson et al. (2001), implying there could be X-ray modulation at the orbital period. {\em Swift} (Gehrels et al. 2004) Target of Opportunity observations were therefore subsequently requested.
Following a short test pointing\footnote{This test was performed in order to determine whether offsetting the pointing to avoid the nearby bright stars (56~Cyg and 57~Cyg) would enable the use of the UVOT.} 
 on 2013 Aug. 10, {\it Swift} monitoring on an orbit-by-orbit basis was performed on 2013 Aug. 14 (Table~\ref{table:obs}), in order to place firmer constraints on the X-ray modulation detected by {\em XMM-Newton}. 14.4~ks of data were collected by both the X-ray (XRT; Burrows et al. 2005) and UV/Optical (UVOT; Roming et al. 2005) telescopes. Since the signal-to-noise of these data was insufficient to draw clear conclusions, further monitoring was performed between 2013 Aug. 23 and 25, collecting an additional 45.9~ks of X-ray data ($\sim$~10~ks in each of the three UVOT UV filters, individual exposures of between $\sim$~100 and 400~s).

 The {\em Swift} observations were performed using a ``slew-in-place'', to ensure that the very bright stars 65 and 57~Cyg were outside the UVOT field of view. The very short snapshots taken before the secondary slew (00032902001, 00032902003, 00032919001, 00032924001 and 00032924003) have not been analysed in this work.  
All the {\em Swift}-XRT data were collected in Photon Counting (PC) mode and 
X-ray data from both {\em XMM-Newton} and {\em Swift} were analysed over 0.3--10~keV, using the standard selection of patterns/grades 0--12 ({\em XMM-Newton}-MOS and {\em Swift}-PC) and 0--4 ({\em XMM-Newton}-pn). For the spectral analysis, times of high background were excluded from the {\em XMM-Newton} data.

\begin{table*}
\caption{Log of the {\em XMM-Newton} and {\em Swift} observations used in this paper. The X-ray and UV columns give the instruments and filters used, respectively. The central wavelengths of the UVOT filters are: $uvw2$~=~1928\AA; $uvm2$~=~2246\AA; $uvw1$~=~2600\AA.} 
\label{table:obs}      
\centering                                      
\begin{tabular}{llllll}          
\hline                       
Satellite & ObsID & Start time UT & X-ray & UV & X-ray exposure (ks) \\
\hline      
{\em XMM} & 0679580201 & 2011 Nov. 24 at 10:32 & pn& --- & 25.6 \\
          &            & 2011 Nov. 24 at 09:34     & MOS1 &  --- &29.3\\
          &            &  2011 Nov. 24 at 09:34    & MOS 2 &  --- &29.3\\
{\em Swift} & 00032902002 & 2013 Aug. 10 at 23:16 & XRT-PC & $w2$,$m2$,$w1$ & 1.1\\
{\em Swift} & 00032902004 & 2013 Aug. 14 at 01:01 & XRT-PC & $w2$ & 14.4\\
{\em Swift} & 00032919002 & 2013 Aug. 23 at 00:54 & XRT-PC & --- & 14.5\\
{\em Swift} & 00032924002 & 2013 Aug. 23 at 20:07  & XRT-PC & $w2$,$m2$,$w1$ & 19.6\\
{\em Swift} & 00032924004 & 2013 Aug. 25 at 00:57& XRT-PC & $w2$,$m2$,$w1$ & 11.8\\

\hline                                             
\end{tabular}
\end{table*}

The AAVSO (American Association of Variable Star Observers) light-curve shows that V751~Cyg has not returned to an optical low state (fainter than $\sim$~magnitude 15) since May 1998; its typical visual magnitude is around 14--14.5. Thus, the {\em XMM-Newton} and {\em Swift} observations presented here were obtained during the (usual) high optical state.

\section{Data Analysis}
\label{anal}

\subsection{Timing}
\label{timinganal}

\begin{figure}
\centering
\includegraphics[width=8cm]{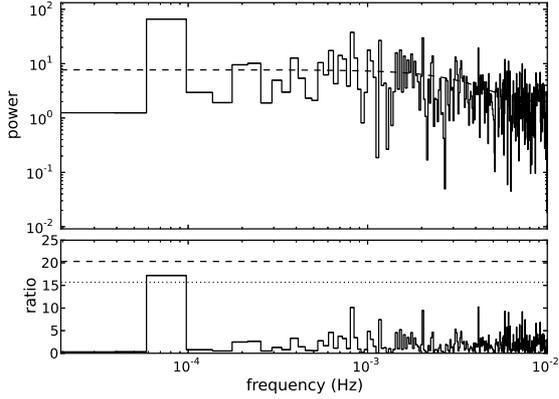}
\caption{Fourier periodogram of the pn 50s light-curve, fitted with a zero-centred Gaussian plus constant model (shown by the dashed line in the top panel). The lower panel shows the 90\% (dotted) and 99\% (dashed) significance levels overplotted on the ratio of twice the data to the model.}
\label{ls}
\end{figure}

Figure~\ref{lc} shows a slow variation with a 12~ks timescale in the combined EPIC (European Photon Imaging Camera; Str{\" u}der et al. 2001, Turner et al. 2001) -- i.e. pn+MOS1+MOS2 -- {\em XMM-Newton} light-curve. 
Periodograms were created using 50~s binned MOS1, MOS2 and pn light-curves (Fig.~\ref{ls} shows the pn results). In each case, the two strongest power spikes occur at periods of 20 min and 3.54~hr (frequencies of 8.3~$\times$10$^{-4}$ and 7.85~$\times$10$^{-5}$~Hz). The longer period (fewer than one and a half full cycles are covered by the complete {\em XMM-Newton} exposure) is consistent with the 3.467~hr modulation noted by Patterson et al. (2001); the 1$\sigma$ uncertainty on our X-ray measurement is $\sim$~0.2~hr, estimated by fitting a sine wave to the light-curve. Following Vaughan (2010), a model consisting of a zero-centred Gaussian (to account for any low frequency noise) and a constant (to account for the Poisson noise) was fitted to the periodogram. The lower panel in Fig.~\ref{ls} shows the ratio of twice the data to the model, with the horizontal lines showing the 90\% (dotted) and 99\% (dashed) detection levels for any period above the red noise. The modulation at the orbital period is thus significant at $>$90\%

In addition, Patterson et al. (2001) mention a shorter period of $\sim$~1000~s, similar to the 20 min one seen in the {\em XMM-Newton} data presented here, but dismiss it as likely due to red noise.  The short period spike in our data is much less than 90\% significant when red noise is included within the power spectral model, thus we do not consider this period further.

\begin{figure}
\centering
\includegraphics[width=6cm, angle=-90]{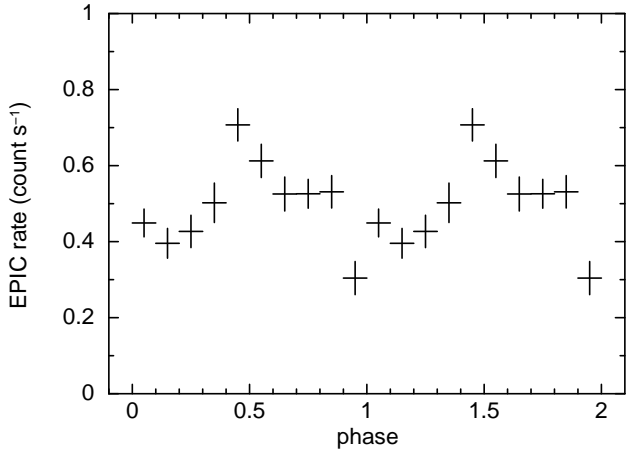}
\caption{EPIC 50s binned light-curves folded into ten phase bins at the orbital period of 3.467~hr.}
\label{xmmfold}
\end{figure}

Figure~\ref{xmmfold} shows the combined EPIC light-curve folded on the Patterson et al. (2001) orbital period of 3.467~hr, using ten phase bins per cycle, with the epoch chosen such that a minimum is observed at phase 1.0. Fitting a constant+sine model to the phase-folded data, the modulation amplitude was measured to be (24~$\pm$~6)\%.

\begin{figure}
\centering
\includegraphics[width=6cm, angle=-90]{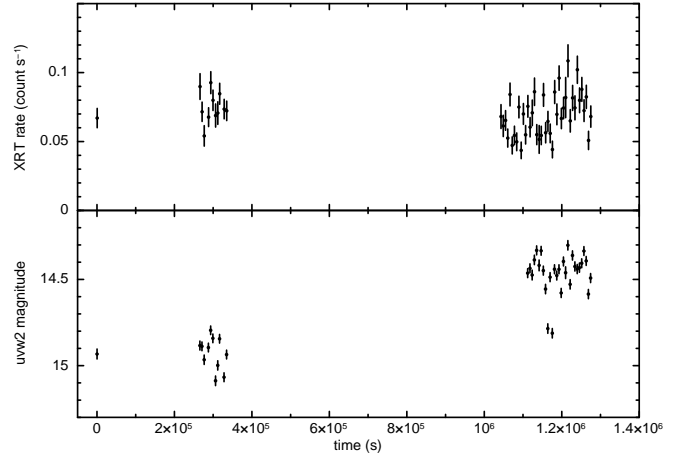}
\caption{{\em Swift} X-ray and $uvw2$ light-curves in seconds since the start of the first {\em Swift} observation: 2013 Aug. 10 at 23:16 UT = MJD 56514.97. V751~Cyg was around half a magnitude brighter in late 2013 August than in mid August, while the X-ray flux remained essentially unchanged. In the text, the observations around 3~$\times$10$^5$~s are referred to as `early', while the data after 10$^6$~s are `late'.}
\label{swiftlc}
\end{figure}

The complete X-ray and $uvw2$ light-curves obtained from {\em Swift} are shown in Figure~\ref{swiftlc}. The XRT data are plotted as one bin per snapshot, the UVOT as one bin per exposure (multiple exposures can occur in a single snapshot, in order to cycle through all three filters). 
The UV brightness of V751~Cyg increased in all filters between 2013 Aug. 10 and 25 by about half a magnitude on average ($uvw1$: $\sim$~mag 14.5--14.0; $uvm2$: $\sim$~mag 14.9--14.4; $uvw2$: $\sim$~mag 15--14.5; the central wavelengths of the filters are given in Table~\ref{table:obs}), while the mean 0.3--10~keV X-ray count rate remained constant at $\sim$~0.07~count s$^{-1}$: the mean count rates for the X-ray data obtained during the 2013 Aug. 14 (hereafter to be referred to as the `early' dataset) and 2013 Aug. 23/25 (`late') epochs are completely consistent with one another.
The two low UVOT points (at 1.16~$\times$~10$^6$ and 1.18~$\times$~10$^6$~s) appear to be real, with no sign of problems in the data; the X-ray flux is not noticeably different at these times. 
A periodogram of the {\em Swift} XRT data (extracted using 50~s bins) does not reveal any significant power spikes; similarly, the UVOT also provides no good independent evidence for a significant period using this method. We note that the barycentric correction for all the data presented here was negligible.

\begin{figure}
\centering
\includegraphics[width=6cm, angle=-90]{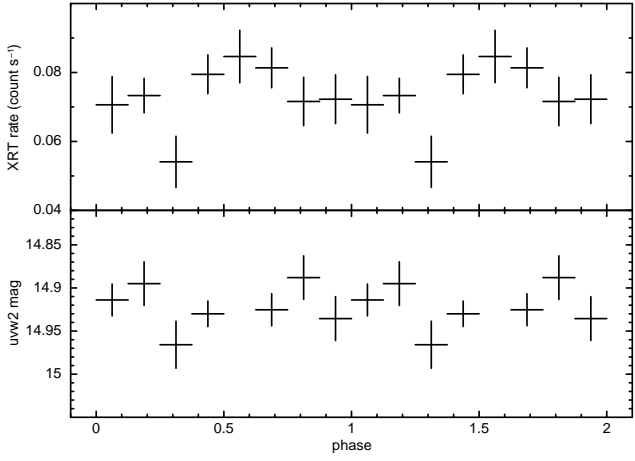}
\caption{{\em Swift} X-ray and UV snapshot light-curves from the 2013 Aug. 14 (`early') monitoring interval folded into eight phase bins at the orbital period of 3.467~hr.}
\label{swiftfoldearly}
\end{figure}

\begin{figure}
\centering
\includegraphics[width=6cm, angle=-90]{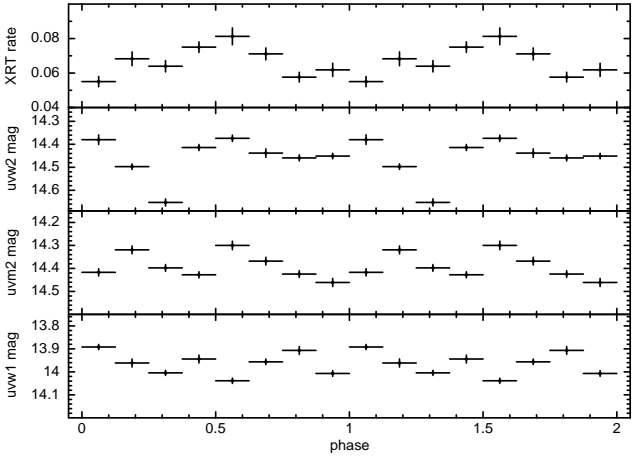}
\caption{ {\em Swift} X-ray and UV snapshot light-curves from the 2013 Aug. 23/25 (`late') monitoring intervals folded into eight phase bins at the orbital period of 3.467~hr. The XRT rate is in units of count~s$^{-1}$.}
\label{swiftfoldlate}
\end{figure}

Figs.~\ref{swiftfoldearly} and \ref{swiftfoldlate} show the {\em Swift} X-ray and UV data from  the `early' and `late' epochs folded on the same 3.467~hr period and epoch as the {\em XMM-Newton} data. While a periodogram may not have shown obvious power at the known period, the X-ray data show the orbital modulation particularly clearly in the `late' dataset, with modulation amplitudes of (9~$\pm$~7)\% and (15~$\pm$~5)\% for the 'early' and 'late' datasets respectively. The amplitude of the modulation in these {\em Swift} data is significantly lower than that seen by {\em XMM-Newton}.
Regarding the UV data, although the eye may not see an obvious modulation, a constant+sine fit to the data is, in all cases, better than a simple constant at the 90\% level, with modulation amplitudes of (0.083~$\pm$~0.082)\% and (0.40~$\pm$~0.08)\% for $uvw2$ ('early' and 'late'), (0.15~$\pm$~0.09)\% for $uvm2$ and (0.26~$\pm$~0.08)\% for $uvw1$ (the $uvm2$ and $uvw1$ filters were not used during the `early' epoch). With the possible exception of the `late' $uvw1$ data, the X-ray and UV modulations are approximately in phase with each other.  All errors are given at the 90\% confidence level.

\begin{figure}
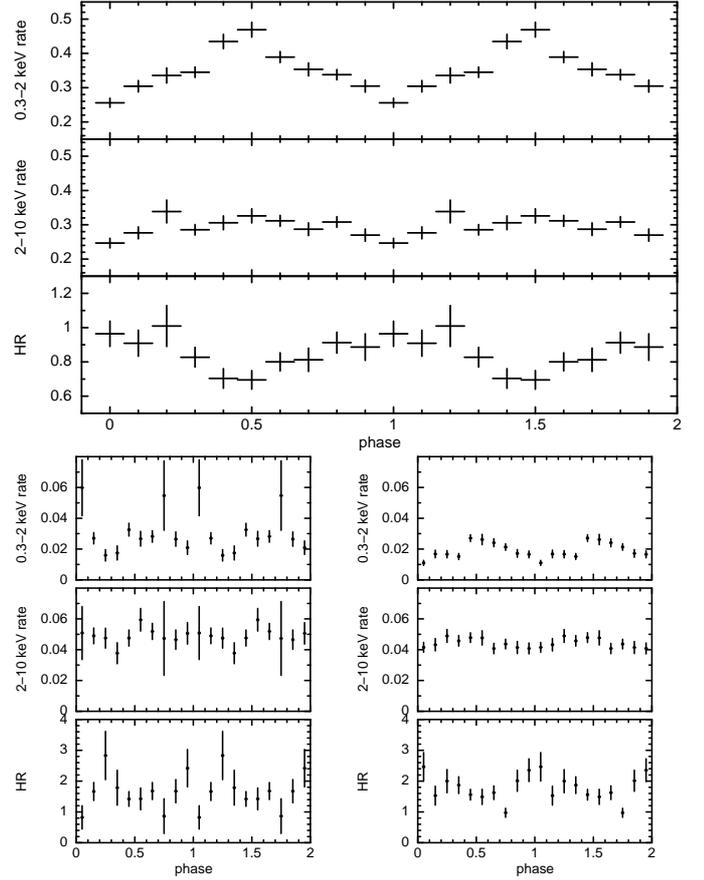

\centering
\includegraphics[width=6cm, angle=-90]{HR_fold_12482_EPIC.ps}
\includegraphics[width=5.6cm, angle=-90]{HR_12482_early.ps}
\includegraphics[width=5.6cm, angle=-90]{HR_12482_late.ps}
\caption{Combined EPIC (top panel) and {\em Swift} `early' and `late' intervals (bottom panel, left and right) soft and hard band light-curves and hardness ratios (HR) folded on the orbital period. The X-ray rates are in units of count~s$^{-1}$ for the instruments in question.}
\label{hr}
\end{figure}

The hardness ratios, calculated as 2--10~keV/0.3--2~keV, were created for the combined EPIC and {\em Swift}-XRT (`early' and `late' intervals) data and folded on the orbital period (see Fig.~\ref{hr}); the EPIC hardness plot is in anti-phase with the count-rate (i.e., the emission is harder when fainter), with a modulation amplitude of (14~$\pm$~6)\%. The `late' {\em Swift} data have a hardness ratio modulation of  (24~$\pm$~12)\%, also being harder when fainter. 

The AAVSO observations were checked for evidence of the orbital period, but the data were too sparse to provide any useful constraints.

\subsection{Spectral fitting}
\label{specanal}

Spectra extracted for each of the {\em XMM-Newton} EPIC instruments were analysed simultaneously using the {\sc xspec} fitting package (version 12.8.0j).
The spectra could be well fitted by two absorbed optically-thin emission components using the {\sc apec} model (Smith et al. 2001), together with an ionised ($\sim$~6.7~keV) iron emission line (Table~\ref{specfit}). The Wilms abundances (Wilms, Allen \& McCray 2000) and photoelectric absorption cross-section from Verner et al. (1996) were used with the T{\" u}bingen-Boulder ({\sc tbabs}) absorption model. The {\sc apec} abundances were left fixed at solar, since they were not constrained by the spectra. Optimising the spectral model using Cash statistics, but with Pearson's $\chi^2$ (weight model) as the test statistic, resulted in a good fit (C-stat/dof = 1477/1760; $\chi^2$/dof = 1667/1760; see Table~\ref{specfit}). The dual temperature model is an improvement over the single component at the $\sim$~3$\sigma$ level using the F-test.

Phase-resolved spectra were also extracted for the pn dataset (since these data have the highest signal-to-noise). Fitting the spectra simultaneously, keeping the temperature of the emission components tied between the two spectra, the absorbing column decreases from (3.9$^{+0.5}_{-0.8}$)~$\times$~10$^{21}$~cm$^{-2}$ to (2.7$^{+0.4}_{-0.6}$)~$\times$~10$^{21}$~cm$^{-2}$ between the 'faint/hard' (phases 0.0--0.25 and 0.75--1.0) and `bright/soft' (phase 0.25--0.75) intervals.

\begin{figure}
\centering
\includegraphics[width=6cm, angle=-90]{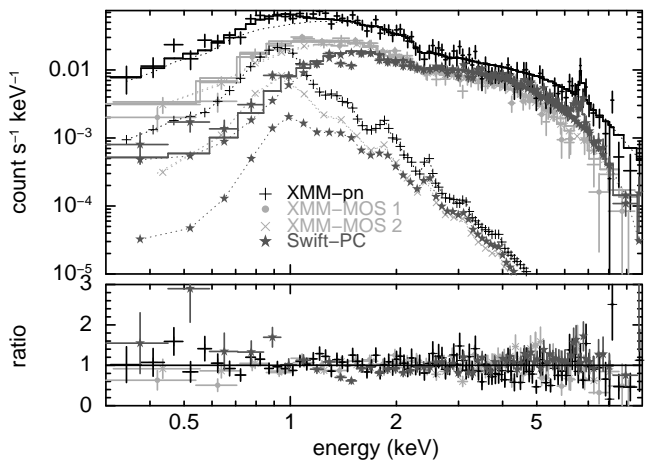}
\caption{{\em XMM-Newton}-EPIC (2011 Nov.) and {\it Swift}-XRT (2013 Aug.) spectra of V751 Cyg, fitted with two absorbed optically thin emission components (the separate components are shown by the fainter dotted lines) and a narrow Gaussian line, as detailed in the text and Table~\ref{specfit}. The model lines are from the fit with the emission parameters tied between the EPIC and XRT spectra; only the absorbing column was allowed to differ.}
\label{spec}
\end{figure}

The {\em Swift} observations occurred 21 months after the {\em XMM-Newton} data were collected. At this time, as shown in Table~\ref{specfit}, the spectra were still better parameterised by a dual temperature fit (C-stat/dof = 641/685; $\chi^2$/dof = 702/685), although the cooler of the two spectral components was cooler still, while the higher temperature component was hotter. The best-fit absorbing column had increased by almost a factor of six (Fig.~\ref{spec}; Table~\ref{specfit}). Fits to spectra extracted from the individual {\em Swift} datasets were consistent with each other, so all data were combined to obtain the best constraints.

Keeping the temperatures of the emission components tied between the two epochs ({\em XMM-Newton} and {\em Swift}) and just allowing the absorbing column to vary, provides an acceptable simultaneous fit, indicating that the main difference is likely to be due to the increase in absorption, with variations in temperature being a secondary effect.
In this fixed temperature case, the emission component temperatures are kT~=~0.94$^{+0.07}_{-0.14}$ and $>$54~keV, with  N$_{\rm H}$~=~(3.4~$\pm$~0.2)~$\times$~10$^{21}$ cm$^{-2}$ ({\em XMM-Newton}) and (11.8$^{+0.6}_{-0.5}$)~$\times$~10$^{21}$ cm$^{-2}$ ({\em Swift}).

Using this joint fit, and a distance to V751~Cyg of 500~pc [as estimated by Greiner et al. (1999) from E(B-V) measurements of 0.25~$\pm$~0.05], the unabsorbed bolometric luminosities are 2.4~$\times$~10$^{32}$~erg~s$^{-1}$ and 6.8~$\times$~10$^{32}$~erg~s$^{-1}$ for the {\em XMM-Newton} and {\em Swift} observational epochs respectively.  We note that the absorption values determined from the fits are significantly in excess of those expected from the E(B-V) value, which corresponds to N$_{\rm H}$~=~(1.5~$\pm$~0.3)~$\times$~10$^{21}$~cm$^{-2}$.

\begin{table*}
\caption{Spectral fitting results for the single and double-temperature models. The sixth column gives the unabsorbed flux over 0.3--10~keV. Cash was used as the fit statistic (i.e. minimisation of the model was done using C-stat), while $\chi^2$ is given as the corresponding test statistic. Errors are at the 90\% confidence level.} 
\label{specfit}      
\centering                                      
\begin{tabular}{llllllll}          
\hline                       
Instrument & kT  & kT & Line energy  & N$_{\rm H}$ & Unabs. flux & C-stat/dof & $\chi^2$/dof \\
&(keV) & (keV) & (keV) & (10$^{21}$ cm$^{-2}$) & (erg cm$^{-2}$ s$^{-1}$)\\
\hline      
{\em XMM}-EPIC &  & 13$^{+3}_{-2}$ & 6.51$^{+0.06}_{-0.05}$& 3.4~$\pm$~0.2  & 2.24~$\times$~10$^{-12}$ & 1488/1762 & 1689/1762\\ 
{\em XMM}-EPIC & 0.76$^{+0.19}_{-0.12}$ & 16$^{+4}_{-3}$ &  6.53$^{+0.14}_{-0.05}$  & 3.8$^{+0.3}_{-0.2}$ & 2.32~$\times$~10$^{-12}$ & 1477/1760 & 1667/1760\\ 
{\em Swift}-XRT & & $>$61 & 6.72$^{+0.05}_{-0.03}$ & 10.4~$\pm$~0.5 & 6.65~$\times$~10$^{-12}$ & 693/687 & 788/687\\
{\em Swift}-XRT & 0.16$^{+0.04}_{-0.01}$ & $>$50 & 6.74$^{+0.03}_{-0.04}$ &  20~$\pm$~2 & 1.07~$\times$~10$^{-10}$ &  641/685 & 702/685\\
\hline
{\em XMM}+{\em Swift} & 0.94$^{+0.07}_{-0.14}$ & $>$54 & 6.67$^{+0.02}_{-0.03}$ & {\em XMM}: 3.4~$\pm$~0.2 &{\em XMM}: 2.45~$\times$~10$^{-12}$  & 2184/2450 & 2429/2450\\
 & & & & {\em Swift}: 11.8$^{+0.6}_{-0.5}$ &{\em Swift}: 6.97~$\times$~10$^{-12}$ \\
\hline                                             
\end{tabular}
\end{table*}

{\em Swift} obtains X-ray and UV data simultaneously because of the co-aligned XRT and UVOT, allowing the creation of spectral energy distributions (SEDs). The best-fit model for the XRT data, with an additional component ({\sc zdust} command within {\sc xspec}) to account for extinction due to dust grains (assuming a Milky Way extinction curve; Pei 1992), under-predicts the UVOT flux by a factor of 10--20, indicating the excess UV emission likely arises from a distinct emission component, perhaps the inner accretion disc.

A simple blackbody model fitted to the fluxes from the three different UV filters provides a very poor result. UV spectra obtained for other nova-like variables are frequently dominated by absorption lines during their high optical states (e.g., VY~Scl, Hamilton \& Sion 2008), which may explain the divergence from a blackbody continuum for the data presented here. 

Since the UVOT data cannot be sensibly modelled with a blackbody, the UV luminosities were instead estimated by multiplying the flux densities at the central wavelength by the FWHM of the filters (given in Poole et al. 2008), and dereddening using a standard Milky Way extinction curve (Pei 1992).
This method provides the band-limited values given in Table~\ref{table:uvlum}. Summing the fluxes from the separate filters leads to lower limit estimates of the total UV flux of 3.7~$\times$~10$^{32}$~erg~s$^{-1}$ during the `early' interval, rising to 6.1~$\times$~10$^{32}$~erg~s$^{-1}$ during the `late' epoch. The 0.3--10~keV band-limited X-ray luminosity during these observations is 2.0~$\times$~10$^{32}$~erg~s$^{-1}$, indicating that most of the accretion luminosity is in the UV band. 

\begin{table}
\caption{Band-limited UVOT luminosities for the `early' and `late' {\em Swift} observations.} 
\label{table:uvlum}      
\centering                                      
\begin{tabular}{lll}          
\hline                       
Filter & `Early' Luminosity. & `Late' Luminosity \\
& (erg~s$^{-1}$) & (erg~s$^{-1}$)\\
\hline      
$uvw2$ & 1.3~$\times$~10$^{32}$ & 2.1~$\times$~10$^{32}$\\
$uvm2$ & 8.9~$\times$~10$^{31}$ & 1.4~$\times$~10$^{32}$\\
$uvw1$ & 1.5~$\times$~10$^{32}$ & 2.6~$\times$~10$^{32}$\\

\hline                                             
\end{tabular}
\end{table}

\section{Discussion}
\label{disc}

We have reported the first X-ray observations of V751 Cyg with
significant spectral resolution; these show that, in the optical high state, the emission
originates in optically thin plasma, as commonly seen in cataclysmic
variables (Mukai 2000; Baskill, Wheatley \& Osborne 2001). Our observations revealed X-ray
flux and spectral modulation at the orbital period, with phase-resolved spectroscopy suggesting that absorption is most likely the cause. We also saw UV flux modulation in some light-curves folded on the orbital period.

Greiner et al. (1999) proposed that V751~Cyg was a super-soft X-ray
source during the optical low state, based on {\em ROSAT}-HRI spectral
analysis. Previous work on other VY~Scl stars using {\em ROSAT} data,
both by Schlegel \& Singh (1995) and Greiner (1998), also suggested
that blackbodies were good fits to their X-ray spectra in the optical
high state. Schlegel \& Singh (1995) analysed two such stars, MV~Lr and
KR~Aur. However, Mauche \& Mukai (2002) analysed {\em ASCA} GIS and SIS spectra from KR~Aur, as well as from another VY~Scl star, TT~Ari, finding that they
were well fitted by optically thin thermal plasma models rather than
blackbodies. Although KR~Aur was in an intermediate optical state at
this time, TT~Ari was in its high optical state, thus raising a question mark regarding the earlier lower spectral resolution reports of
blackbody emission from this class of object. Extending the result of Mauche \& Mukai (2002), Zemko et al. (2014) found that {\em Swift} X-ray spectral
fits are clearly consistent with a collisionally ionised optically thin plasma, rather than a blackbody,
when considering four VY~Scl stars observed in both high and low optical states. Balman, Godon \& Sion
(2014) analysed X-ray datasets for two VY Scl systems, some of which were also used by Zemko et al. (2014), and
a novalike CV in their optical high-state, again finding that optically thin plasma emission provided the
best fit. Our
results presented here for V751~Cyg are in agreement with the 21st century
findings for other VY~Scl stars: there is no sign of blackbody emission in the {\em
XMM-Newton} and {\em Swift} observations of V751~Cyg, at least in the
high optical state.

Since Greiner et al. (1999) found an X-ray detection during the
optical low state but not during the high state of V751~Cyg, they
suggested that there was an anti-correlation between the X-ray and
optical intensities. The upper limit Greiner (1998) derived from the
non-detection by the PSPC during optical high states of V751~Cyg (L$_{\rm
X}$~$<$~2--6~$\times$~10$^{30}$~erg~s$^{-1}$) is lower than
the optical high state, observed, 0.1--2.4~keV luminosity we estimate here: $\sim$~1.4--2~$\times$~10$^{31}$~erg~s$^{-1}$, though of the same order, and two orders of magnitude
less than the luminosity given by Greiner et al. (1999) for the low
state observation. Thus, our data are not inconsistent with there
being an anti-correlation between the different bands. Clearly a modern X-ray observation of V751~Cyg in a low optical state is needed to verify this supposed inverse relationship. In their investigation of VY~Scl stars, however,
Zemko et al. (2014) found that, although BZ~Cam showed a larger X-ray
flux during the optical low state, the other three sources in their
sample (MV~Lyr, TT~Ari and V794~Aql) did not reveal such an
anti-correlation.

The X-ray and UV data of V751~Cyg are both modulated, in phase with one another, at the orbital period determined by Patterson et al. (2001). This suggests that the X-ray and UV emission come from the same general area which is partially blocked from our view during during each orbit. HV~Cet (Beardmore et al. 2012) and Nova~Mon~2012 (Page et al. 2013) showed a similar pattern [also seen in compact binary supersoft sources, as discussed in Beardmore et al. (2012)], which was proposed to be due to a raised portion of the accretion disc rim blocking the emission site. 
It has previously been suggested that V751~Cyg could be an SW~Sex star (Zellem et al. 2009), because of the similarities in orbital period, permanent superhumps and drops into low states. Knigge et al. (2000) observed the SW~Sex type star DW~Uma, finding that it likely has an optically thick accretion disc rim blocking the view of the WD during its high state. Such a flared disc, thicker at the disk-stream
impact region, could also lead to the X-ray/UV modulation seen here in V751~Cyg.  Patterson et al. (2001) found binary phase-dependant blue-shifted H$\alpha$ absorption in the optical spectra
of V751~Cyg, which they ascribed to an azimuthally asymmetric wind; this seems likely to be related to the
X-ray phase-dependent absorption we have observed.

\section{Summary}

A serendipitous X-ray observation of the anti-dwarf nova V751~Cyg was
identified in the 3XMM-DR4 catalogue.  This, together with more recent
observations by {\em Swift}, also taken in the optical high state,
clearly show, for the first time in the X-ray band, optically thin
emission modulated at the orbital period of 3.467~hr noted by Walker
\& Bell (1980) and Patterson et al. (2001); this modulation was also
seen in phase in the UV. Our high-optical state X-ray luminosity is
greater than the upper limit of Greiner et al. (1999), but does not
rule out their hypothesis of anti-correlated optical and X-ray
luminosities (although this is ruled out for some similar objects by other recent observations). Early reports of blackbody emission from
some VY Scl stars in their high states are not confirmed in the case of
V751 Cyg. Our likely discovery of binary phase-dependant X-ray
absorption, together with the observation of phase-dependant wind
absorption in H$\alpha$ by Patterson et al. (2001), suggests a locally
raised accretion disk rim at the disk-stream interaction region as
the source of both effects.

\begin{acknowledgements}

KLP, JPO, APB and PAE acknowledge support from the UK Space
Agency. This work made use of data supplied by the UK Swift Science
Data Centre at the University of Leicester.  We thank the Swift MOC
team for their support of these difficult observations, made possible
by the careful avoidance of nearby very bright stars.

\end{acknowledgements}

\end{document}